\documentclass[twocolumn,prb,showpacs,floatfix]{revtex4}

\usepackage{graphicx}
\usepackage{bm}
\usepackage{color}
\usepackage{amsmath}
\usepackage{amsfonts}
\usepackage{amssymb}

\begin{document}


\title{Magnetic phase transitions in the two-dimensional frustrated quantum antiferromagnet Cs$_2$CuCl$_4$}
\author{Y. Tokiwa$^1$}
\altaffiliation[Present address: ]{Los Alamos National Laboratory, Los Alamos, New Mexico 87545, USA}
\author{T. Radu$^1$}
\author{R. Coldea$^2$}
\author{H. Wilhelm$^1$}
\author{Z. Tylczynski$^3$}
\author{F. Steglich$^1$}
\affiliation{$^1$Max-Planck Institute for Chemical Physics of Solids, D-01187 Dresden, Germany}
\affiliation{$^2$Oxford Physics, Clarendon Laboratory, Parks Road, Oxford OX1 3PU, United Kingdom}
\affiliation{$^3$Institute of Physics, Adam Mickiewicz University, Umultowska 85, 61-614 Poznan,
Poland}

\date{\today}

\begin{abstract}
We report magnetization and specific heat measurements in the 2D
frustrated spin-1/2 Heisenberg antiferromagnet Cs$_2$CuCl$_4$ at
temperatures down to 0.05\ K and high magnetic fields up to 11.5\
T applied along $a$, $b$ and $c$-axes. The low-field
susceptibility $\chi (T)\simeq M/B$ shows a broad maximum around
2.8\,K characteristic of short-range antiferromagnetic
correlations and the overall temperature dependence is well
described by high temperature series expansion calculations for
the partially frustrated triangular lattice with $J$=4.46\,K and
$J'/J$=1/3. At much lower temperatures ($\leq 0.4$ K) and in
in-plane field (along $b$ and $c$ -axes) several new
intermediate-field ordered phases are observed in-between the
low-field incommensurate spiral and the high-field saturated
ferromagnetic state. The ground state energy extracted from the
magnetization curve shows strong zero-point quantum fluctuations
in the ground state at low and intermediate fields.
\end{abstract}


\maketitle

\section{Introduction}

Cs$_2$CuCl$_4$ is a quasi-2D Heisenberg antiferromagnet with $S$=1/2 Cu$^{2+}$ spins arranged in a triangular lattice with spatially-anisotropic
couplings.\cite{Coldea01} The weak interlayer couplings stabilize magnetic order at temperatures below 0.62 K into an incommensurate spin spiral. The
ordering wavevector is largely renormalized from the classical large-$S$ value and this is attributed to the presence of strong quantum fluctuations
enhanced by the low spin, geometric frustrations and low dimensionality.\cite{renormalization,Veillette} The purpose of the thermodynamic
measurements reported here is to probe the phase diagrams in applied magnetic field and see how the ground state spin order evolves from the
low-field region, dominated by strong quantum fluctuations, up to the saturated ferromagnetic phase, where quantum fluctuations are entirely
suppressed by the field. Intermediate fields are particularly interesting as the combination of (still) strong quantum fluctuations, potentially
degenerate states due to frustration and an effective ``cancelling'' of small anisotropies by the applied field may stabilize non-trivial forms of
magnetic order.

The Hamiltonian of Cs$_2$CuCl$_4$ has been determined from measurements of the magnon dispersion in the saturated ferromagnetic phase.\cite{Coldea02}
The exchanges form a triangular lattice with spatially-anisotropic couplings as shown in Fig.\ 1(b) with exchanges $J=0.374(5)$ meV (4.34~K) along
$b$, $J^{\prime}=0.34(3)J$ along the zig-zag bonds in the $bc$ plane, and weak interlayer couplings $J^{\prime\prime}=0.045(5) J$ along $a$. In
addition there is also a small Dzyaloshinskii-Moriya interaction $D=0.053(5) J$, which creates an easy-plane anisotropy in the ($bc$) plane (for
details see Ref. \onlinecite{Coldea02}). Neutron diffraction measurements have shown rather different behavior depending on the field direction with
respect to the easy-plane. For perpendicular fields (along $a$-axis) incommensurate cone order with spins precessing around the field axis is stable
up to ferromagnetic saturation, however for fields applied along the $c$-axis (in-plane) the incommensurate order is suppressed by rather low fields,
2.1~T compared to the saturation field of 8.0~T along this axis.\cite{Coldea01} The purpose of the present magnetization and specific heat
measurements is to explore in detail the phase diagram in this region of intermediate to high fields. From anomalies in the thermodynamic quantities
we observe that for in-plane field several phases occur in-between the low-field spiral and the saturated ferromagnetic states. From the
magnetization curve we extract the work required to fully saturate the spins and from this we derive the total ground state energy in magnetic field
and the component due to zero-point quantum fluctuations.

\section{Experimental details}

DC-magnetization of a high-quality single crystal of
Cs$_2$CuCl$_4$ grown from solution was measured at temperatures
down to 0.05\,K and high fields up to 11.5\,T using a
high-resolution capacitive Faraday magnetometer\cite{Sak94}. A
commercial superconducting quantum interference device
magnetometer (Quantum Design MPMS) was used to measure the
magnetization from 2\,K to 300\ K. The specific heat measurements
were carried out at temperatures down to 0.05~K in magnetic fields
up to 11.5~T using the compensated quasi-adiabatic heat pulse
method~\cite{Wil04}.

\section{Measurements and results}

\subsection{Temperature-dependence of susceptibility}

\begin{figure}[tbh]
\includegraphics[height=10cm,keepaspectratio]{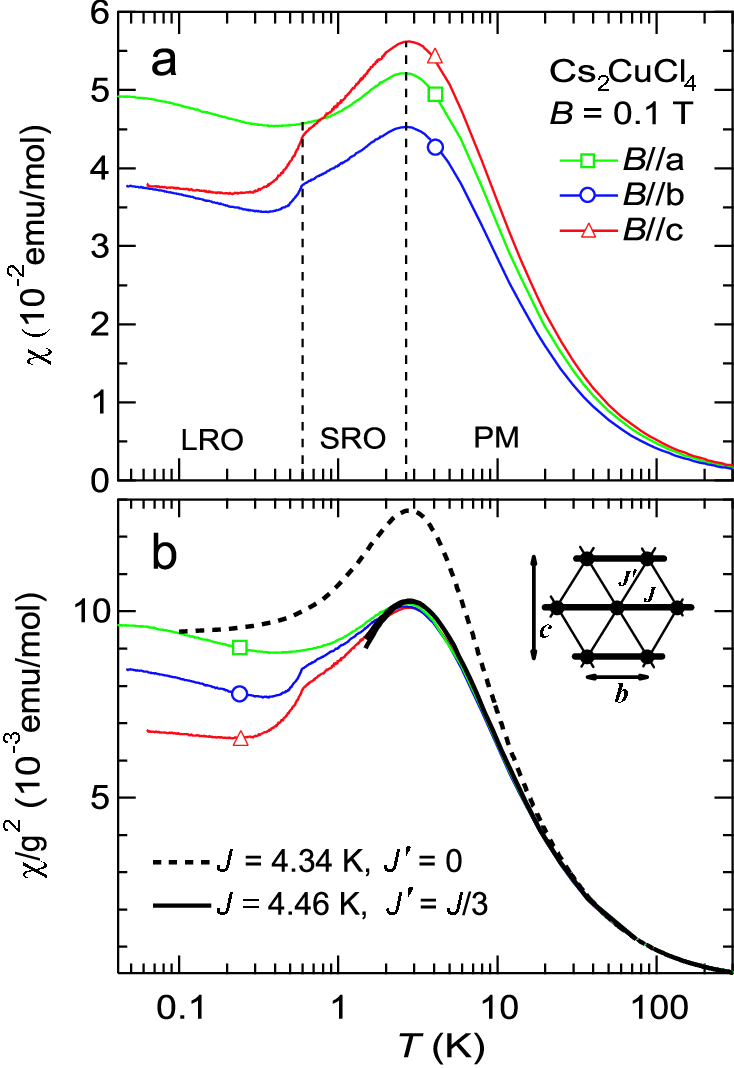}
\caption{(color online)\label{chi}(a) Temperature dependence of
the susceptibility $\chi \simeq M/B$ of Cs$_2$CuCl$_4$ along the
three crystallographic axes. Labels indicate magnetic long range
order (LRO), short-range order (SRO) and paramagnetic (PM). (b)
Susceptibility divided by the $g$-factor squared compared to
calculations for a 2D anisotropic triangular lattice (see inset)
with $J^{\prime}/J=1/3$ and $J=4.46$\,K (thick solid line), and
non-interacting 1D chains with $J^{\prime}=0$ and $J=4.34$\,K
(thick dashed line).}
\end{figure}

\begin{figure}[t]
\includegraphics[width=8cm,keepaspectratio]{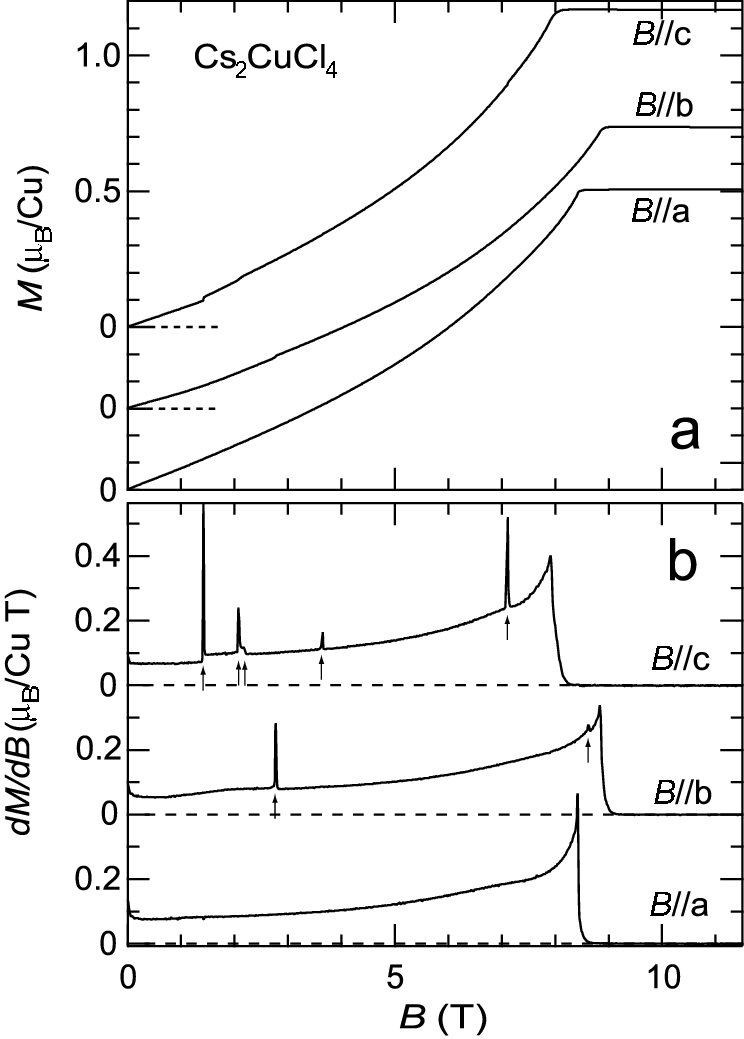}
\caption{\label{M}(a) Magnetization curves of Cs$_2$CuCl$_4$ measured at the base temperature for the field applied along the three crystallographic
axes. The curves for $B\parallel b$ and $c$ are shifted by 0.3 and 0.6\,$\mu_{\rm B}/$Cu, respectively. (b) Susceptibility ${dM}/{dB}$ vs field.
Vertical arrows indicate anomalies associated with phase transitions (see text).}
\end{figure}

We first discuss the temperature-dependence of the magnetic
susceptibility at low field and compare with theoretical
predictions for an anisotropic triangular lattice as appropriate
for Cs$_2$CuCl$_4$. Figure~\ref{chi}(a) shows the measured
susceptibility $\chi\simeq M/B$ in a field of 0.1 T. A Curie-Weiss
local-moment behavior is observed at high temperatures and a broad
maximum, characteristic of short-range antiferromagnetic
correlations, occurs around $T_{\rm max}=2.8(1)$\ K, in agreement
with earlier data.\cite{Carlin} Upon further cooling the $b$- and
$c$-axes susceptibilities show a clear kink at $T_{\rm N}$=0.62\
K, indicating the transition to long-range magnetic order. No
clear anomaly at $T_{\rm N}$ is observed for $B\parallel a$. This
is because the magnetic structure has ordered moments spiralling
in a plane which makes a very small angle ($\sim 17\deg$) with the
$bc$ plane.\cite{Coldea96} In this case the near out-of-plane
($a$-axis) susceptibility is much less sensitive to the onset of
magnetic order compared to the in-plane susceptibility (along $b$
and $c$). Fitting the high-temperature data ($T\geq 20$\ K) to a
Curie-Weiss form $\chi(T)=C/(T+\Theta)$ with
$C=N_{\rm{A}}g^2\mu^2_BS(S+1)/3k_B$ gives $\Theta=4.0 \pm $0.2\ K
and $g$-factors $g_a$=2.27, $g_b$=2.11 and $g_c$=2.36 for the
$a$-, $b$- and $c$-axes, respectively. The $g$-factors are in good
agreement with the values obtained by low-temperature ESR
measurements $g$=(2.20, 2.08, 2.30)\cite{Bailleul94}. When the
susceptibility is scaled by the determined $g$-values, $\chi/g^2$,
the data along all three crystallographic directions overlap
within experimental accuracy onto a common curve for temperatures
above the peak, indicating that the small anisotropy term in the
Hamiltonian (estimated to $\sim$5\% $J$) are only relevant at much
lower temperatures. In the temperature range $T \geq T_{max}$ we
compare the data with high-temperature series expansion
calculations\cite{Weihong04} for a 2D spin-1/2 Hamiltonian on an
anisotropic triangular lattice (see Fig.~\ref{chi}(b) inset). Very
good agreement is found for exchange couplings $J^{\prime}/J=1/3$
and $J$=4.46\,K (0.384 meV) (solid line in Fig.~\ref{chi}(b)). In
contrast, the data departs significantly from the expected
Bonner-Fisher curve for one-dimensional chains ($J^{\prime}=0$ and
$J=4.34$ K).\cite{Weihong04}

\subsection{Magnetization curve and ground-state energy}

Figure~\ref{M} shows the magnetization $M(B)$ and its derivative
$\chi=dM/dB$ as a function of applied field at a base temperature
of 0.05 K for the $a$- and $b$-axes and 0.07 K for the $c$-axis.
For all three axes the magnetization increases linearly at low
field but has a clear overall convex shape and saturates above a
critical field $B_{\rm sat}$=8.44(2), 8.89(2) and 8.00(2)\ T along
the $a$-, $b$- and $c$-axis, respectively. When normalized by the
$g$-values the saturation fields are the same within 2\% for the
three directions, the difference being the same order of magnitude
as the relative strength 5\% of the anisotropy terms in the
Hamiltonian.\cite{Coldea02} The saturation magnetizations $M_{\rm
sat}/g=\langle S_z \rangle$ are obtained to be only 1-2.5\% below
the full spin value of 1/2, which might be due to experimental
uncertainties in the absolute units conversion or a slight
overestimate of the $g$-values by this amount. Including such a
small uncertainty in the $g$-values has only a small effect on the
normalized susceptibility $\chi/g^2$ in Fig.~\ref{chi}(b) and does
not change significantly the results of the comparison with the
series expansion calculation for the anisotropic triangular
lattice.

\begin{figure}[t]
\includegraphics[width=8cm,keepaspectratio]{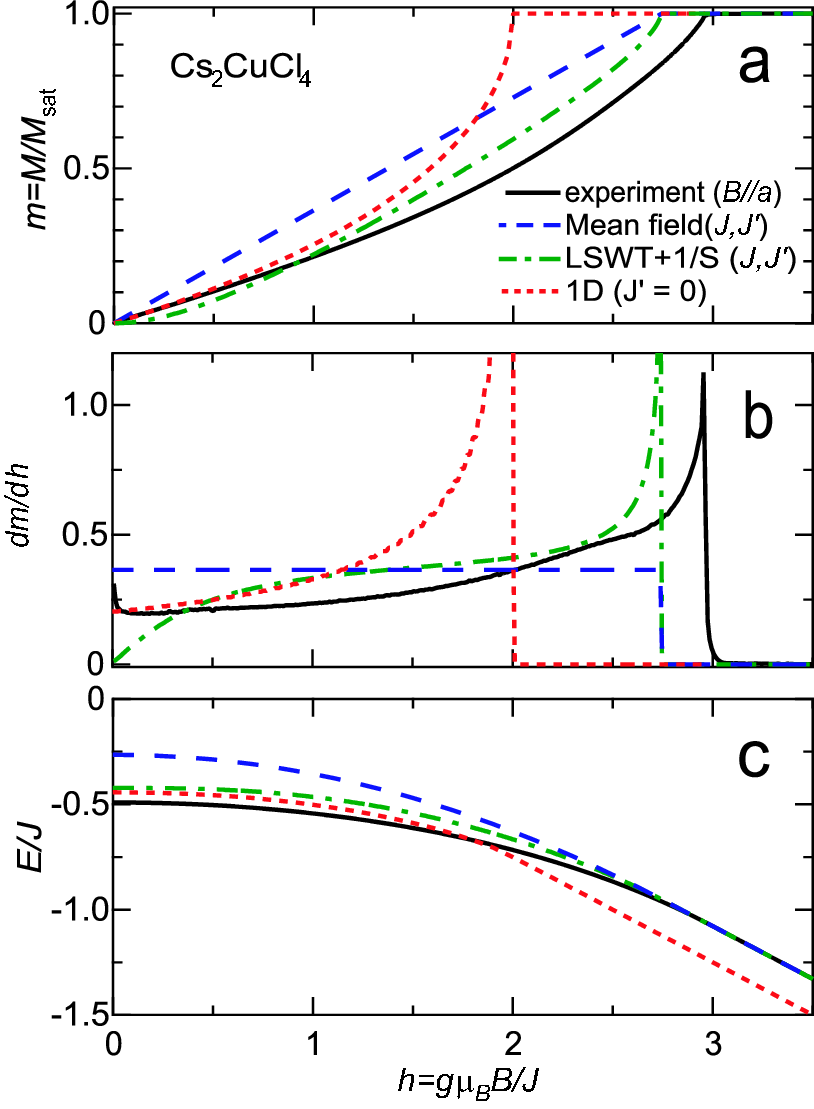}
\caption{(color online)\label{EnergyAnal}(a)Reduced magnetization, (b) susceptibility, and (c) ground-state energy, vs reduced field. Black (solid),
blue (dashed), green (dash-dotted) and red (dotted) lines show experimental data for $B\parallel$a, semiclassical mean-field prediction, linear
spin-wave theory including 1st order quantum corrections and Bethe-ansatz prediction for 1D chains $J^{\prime}=0$, respectively.}
\end{figure}

Before analyzing in detail the various transitions in field
identified by anomalies in the susceptibility $\chi=dM/dB$
(vertical arrows in Fig.\ \ref{M}b) we briefly discuss how the
ground-state energy varies with the applied field, as this gives
important information about the effects of quantum fluctuations.
The ground state energy is obtained by direct integration of the
magnetization curve, i.e.
\begin{equation}
E(B)=E(0)-\int_{0}^{B} M(B) dB
\end{equation}
where the energy (per spin) above the saturation field takes the
classical value $E(B>B_{\rm sat})=J(0)S^2-g \mu_B BS$ because the
ferromagnetic state is an exact eigenstate of the Hamiltonian with
no fluctuations.\cite{Coldea02} Here
$J(0)=\frac{1}{2}\sum_{\delta}J_{\delta}$ is the sum of all
exchange interactions equal to $J+2J^{\prime}$ for the main
Hamiltonian in Cs$_2$CuCl$_4$ [see Fig.\ \ref{chi}b) inset].
Figure~\ref{EnergyAnal} shows comparisons between the experimental
data for $B$$\parallel$$a$ (black solid lines, similar results
obtained using $b$- or $c$-axis data), a mean field calculation
(blue dashed lines), a linear spin-wave theory(LSWT) with 1st
order quantum correction (green dash-dotted lines) and
Bethe-ansatz prediction for 1D chains $J^{\prime}=0$ (red dotted
lines). In magnetic field, a cone structure is predicted by the
classical mean field calculation~\cite{Veillette} $E_{cl}(B<B_{\rm
sat})=S^2\left[J(Q)\cos^{2}\theta+J(0)\sin^2\theta\right]-g\mu_B B
S \sin\theta$ (blue dashed lines) where $Q$ is the classical
ordering wavevector $Q=\cos^{-1}[-J^{\prime}/(2J)]$,
$\theta=\sin^{-1}(B/B_{\rm sat})$, the saturation field is
$g\mu_BB_{\rm sat}=2S[J(0)-J(Q)]$ and
$J(Q)=J\cos(2Q)+2J^{\prime}\cos(Q)$. Here we use $J=0.374$ meV and
$J^{\prime}/J=0.34$ for the main Hamiltonian in Cs$_2$CuCl$_4$.
The experimentally-determined ground state energy (black solid
line) is lower than the classical value (blue dashed line) due to
zero-point quantum fluctuations. The energy difference in zero
field is 85\% of the expected classical energy
$E_{cl}(B=0)=J(Q)S^2$, indicating rather strong quantum
fluctuations in the ground state. The strongly non-linear (convex)
shape of the magnetization curve compared to the
classically-expected linear form $M/M_{\rm sat}(B<B_{\rm
sat})=B/B_{\rm sat}$ [see Fig.~\ref{EnergyAnal}(a)] is a direct
indication of the importance of zero-point quantum fluctuations.
Including 1st order quantum correction to the classical result in
a linear spin-wave approach gives\cite{Veillette}
$E_{LSWT+1/S}=S(S+1)\left[J(Q)\cos^2\theta+
J(0)\sin^2\theta\right]-g\mu_BB(S+1/2)\sin\theta+\langle
\omega_{\bm{k}} \rangle /2$ where $\langle \omega_{\bm{k}} \rangle
$ is the average magnon energy in the 2D Brillouin zone of the
triangular lattice. This improves the agreement with the data
(green dash-dotted lines). Particularly at high fields it captures
better the divergence of the susceptibility [see
Fig.~\ref{EnergyAnal}(b)] at the transition to saturation. The
saturation field is underestimated slightly because we have here
neglected the weak inter-layer couplings $J^{\prime\prime}$=4.5\%
$J$ and the DM interaction $D$=5.3\%J, both of which increase the
field required to ferromagnetically-align the spins. It is also
illuminating to contrast the data with a model of non-interacting
chains ($J^{\prime}=0$, red dotted lines). This would largely (by
48\%) underestimate the observed saturation field and would
predict a rather different functional form for the magnetization
$M_{1D}(B<B_{\rm sat})=M_{\rm
sat}\frac{2}{\pi}\sin^{-1}\left(1-\frac{\pi}{2} +\frac{\pi
J}{g\mu_B B} \right)^{-1}, B_{\rm sat}=2J/g\mu_B$ and
susceptibility ${\partial M_{1D}}/{\partial B}$ compared to the
data, indicating that the 2D frustrated couplings are important.

\begin{figure}[!t]
\includegraphics[width=8cm,keepaspectratio]{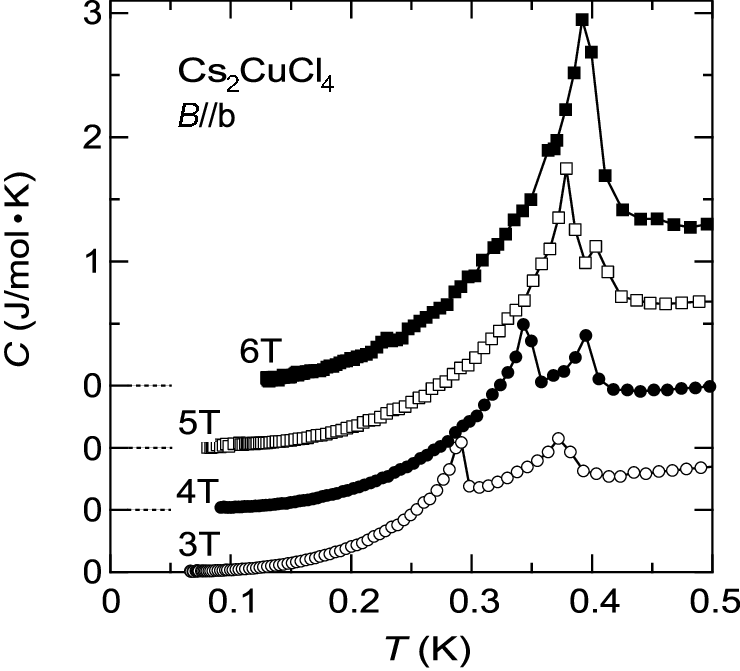}
\caption{\label{C_b} Specific heat as a function of temperature in magnetic fields along $b$-axis. Specific heat data in fields 4, 5 and 6\,T are
shifted upwards by 0.5, 1.0 and 1.5 J/mol$\cdot$K$^2$, respectively.}
\end{figure}

\subsection{Phase diagrams in in-plane field}

When the magnetic field is applied along the $a$-axis
perpendicular to the plane of the zero-field spiral the ordered
spins cant towards the field axis and at the same time maintain a
spiral rotation in the $bc$ plane thus forming a cone. The cone
angle closes continuously at the transition to saturation and as
expected in this case the susceptibility $dM/dB$ observes a sharp
peak followed by a sudden drop as the field crosses the cone to
saturated ferromagnet transition, see Fig.\ \ref{M}(b). However
for fields applied along the $b$- and $c$-axes several additional
anomalies are present in the magnetization curve apart from the
sharp drop in susceptibility upon reaching saturation, indicating
several different phases stabilized at intermediate field.

Before discussing in detail the experimental phase diagrams we
note that for all field directions the magnetization increases in
field up to saturation and no intermediate-field plateaus are
observed, in contrast to the isostructural material
Cs$_2$CuBr$_4$, where a narrow plateau phase occurs for in-plane
field when the magnetization is near $1/3^{\rm rd}$ of
saturation.\cite{Ono04} Such a plateau phase is expected for the
fully-frustrated ($J^{\prime}/J=1$) triangular antiferromagnet and
originates from the formation of the gapped collinear up-up-down
state in field. The absence of a plateau in Cs$_2$CuCl$_4$ is
probably related to the weaker frustration
($J^{\prime}/J$=0.34(3)) compared to
Cs$_2$CuBr$_4$($J^{\prime}/J\sim 0.5$)~\cite{Ono04}.

\begin{figure}[!t]
\vspace{0.5cm}
\includegraphics[width=8cm,keepaspectratio]{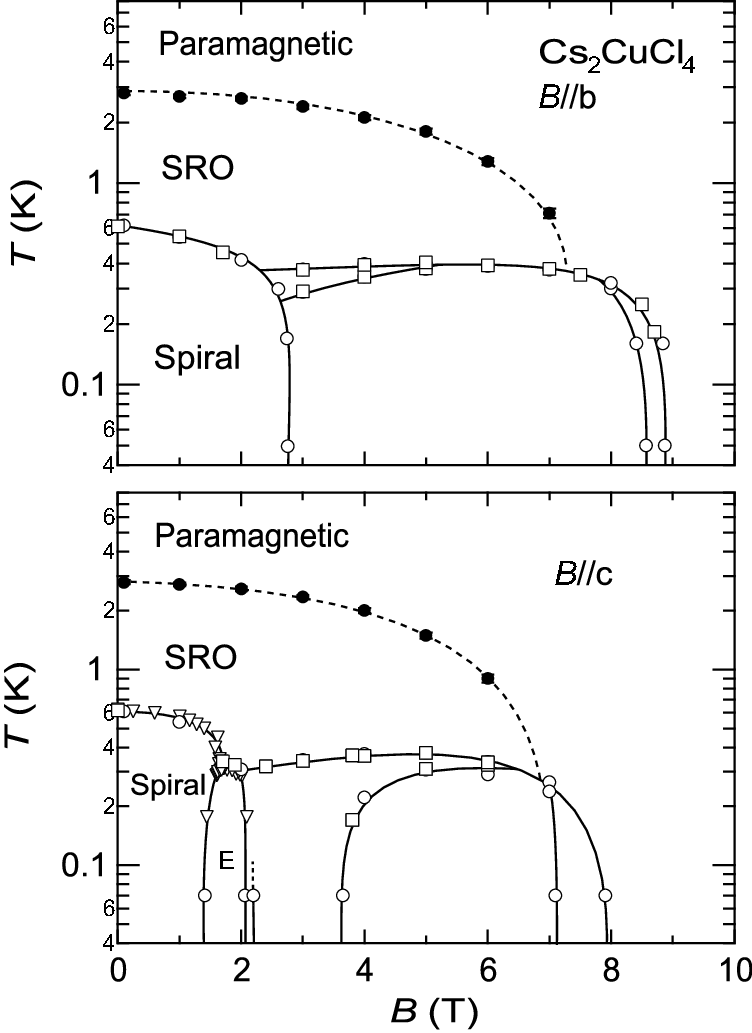}
\caption{\label{B-T} $B-T$ phase diagrams of Cs$_2$CuCl$_4$ for $B\parallel$ $b$- and $c$-axis. Data points of open circles (magnetization), squares
(specific heat) and triangles (neutrons~\cite{Coldea01}) connected by solid lines indicate phase boundaries. Solid circles show positions of the
maximum in the temperature dependence of the magnetization and indicate a cross-over from paramagnetic to short-range order(SRO). "E" on the phase
diagram for $B\parallel c$-axis denotes the elliptical phase.~\cite{Coldea01}}
\end{figure}

A difference in the phase diagrams in field applied along the
$a$-axis and in the $bc$ plane in Cs$_2$CuCl$_4$ is expected based
on the presence of small DM terms in the spin Hamiltonian, which
create a weak easy-plane anisotropy in the $bc$
plane.\cite{Coldea02} Semi-classical calculations which take this
anisotropy into account predict two phases below
saturation:\cite{Veillette} a distorted spiral at low field
separated by a spin-flop like transition from a cone at
intermediate field. The data in Fig.\ \ref{M}(b) however observe
more complex behavior with several different intermediate-field
ordered phases. Also early neutron scattering measurements did not
observe the characteristic incommensurate magnetic Bragg peaks
expected for a cone structure at $B>$2.1 T
$\parallel$$c$-axis~\cite{Coldea01}, suggesting that the magnetic
structure at intermediate field may be quite different from the
classical prediction and may be stabilized by quantum fluctuations
beyond the mean-field level. To map out the extent of the various
phases in in-plane field we have made a detailed survey of the
$B-T$ phase diagram using both temperature and field scans in
magnetization and specific heat and the resulting phase diagrams
are shown in Fig.\ \ref{B-T}. Below we describe in detail the
signature of those transitions in specific heat and magnetization
data, first for field along the $b$-axis, then $c$-axis.

Magnetization and differential susceptibility $dM/dB$ in field
along $b$ are shown in Fig.~\ref{M}. $dM/dB$ shows a sharp peak at
$B$=2.76\,T and an additional small peak at $B$=8.57\,T and those
two anomalies indicate two new phases at base temperature below
the saturation field and above the spiral phase. To probe the
extent in temperature of those phases we show in Fig.\ \ref{C_b}
specific heat measurements as a function of temperature at
constant magnetic field. At 3\ T two peaks are clearly observed
indicating two successive phase transitions upon cooling from high
temperatures. The lower critical temperature increases rapidly
with increasing field and gradually approaches the upper
transition at 5\ T and the two peaks appear to merge at 6\ T.

\begin{figure}[!t]
\includegraphics[width=8.5cm,keepaspectratio]{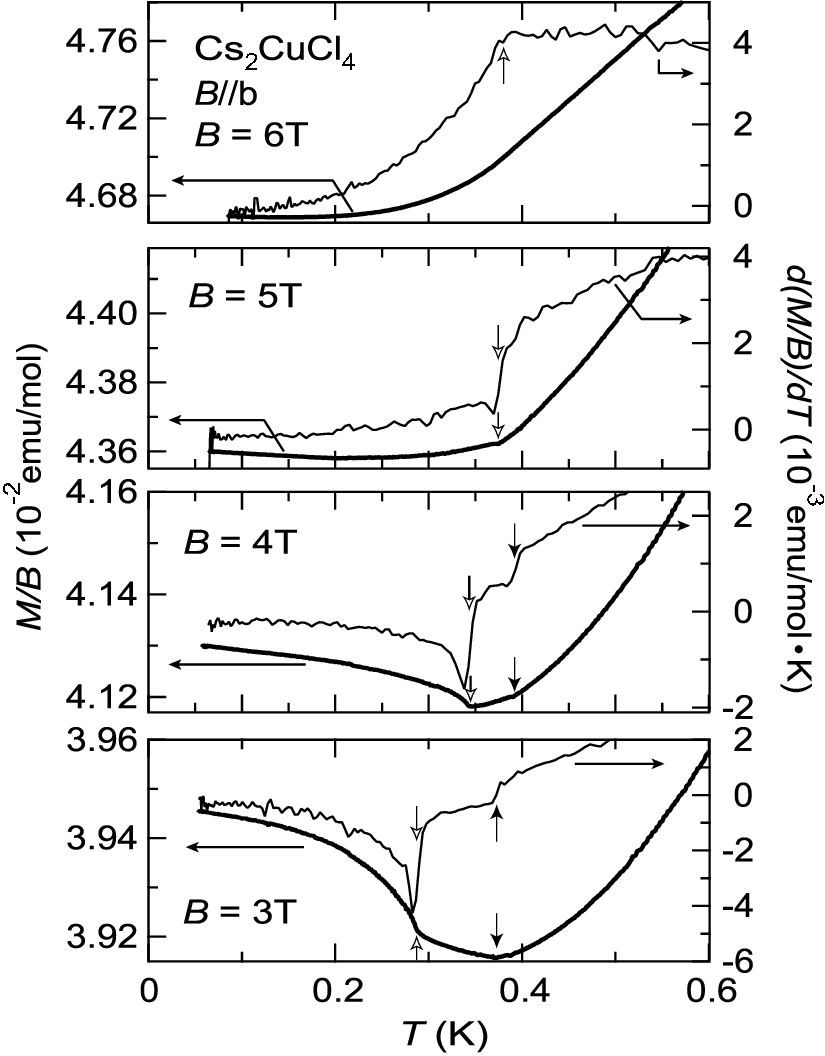}
\caption{\label{M(T)b} Magnetization normalized by applied field $M/B$ (thick solid lines, left axis) and its derivative $d(M/B)/dT$ (thin solid
lines, right axis) as a function of temperature for $B\parallel b$. Vertical arrows indicate anomalies.}
\end{figure}

\begin{figure}[!ht]
\includegraphics[width=8cm,keepaspectratio]{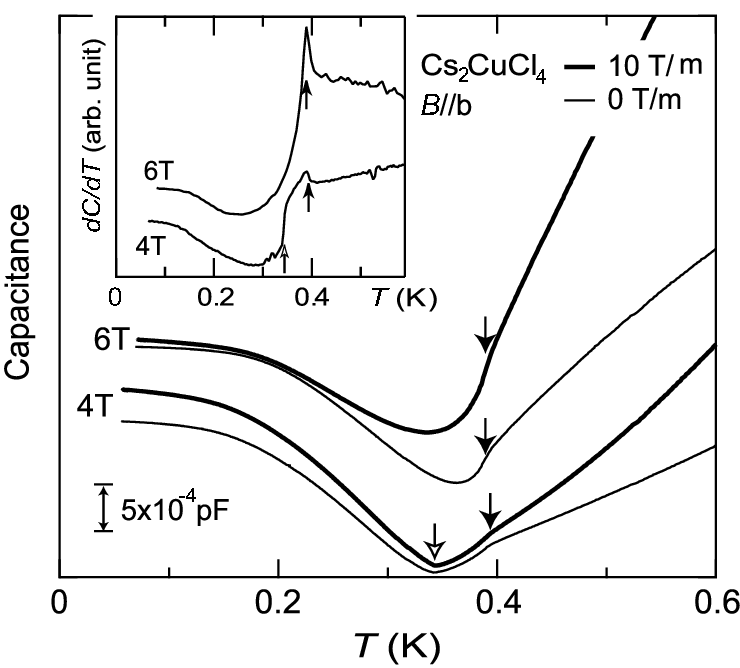}
\caption{\label{torque} Raw capacitance data as a function of temperature in magnetic field of 4 and 6 T applied along $b$-axis. Data are vertically
shifted for clarity. Filled and open arrows indicate anomalies. Thick (thin) solid lines correspond to measurements with (without) gradient field.
Inset shows the temperature derivative of capacitance data in gradient field 10T/m. The curves are shifted vertically for clarity.}
\end{figure}

\begin{figure}[ht]
\includegraphics[width=8cm,keepaspectratio]{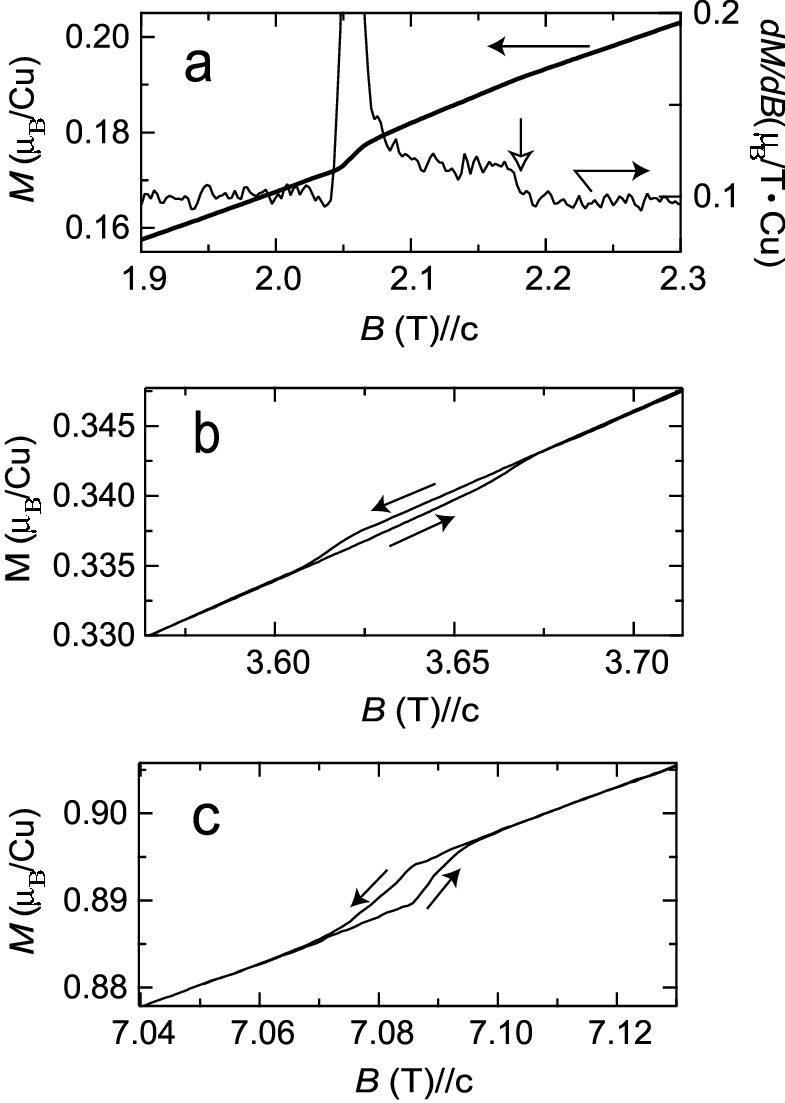}
\caption{\label{AnormM(B)C} Expanded plots of magnetization and susceptibility $\chi=dM/dB$ as a function of field along $c$-axis. Data are identical
to those from Fig.~2(a) and (b).}
\end{figure}

\begin{figure}[!tb]
\includegraphics[width=7.5cm,keepaspectratio]{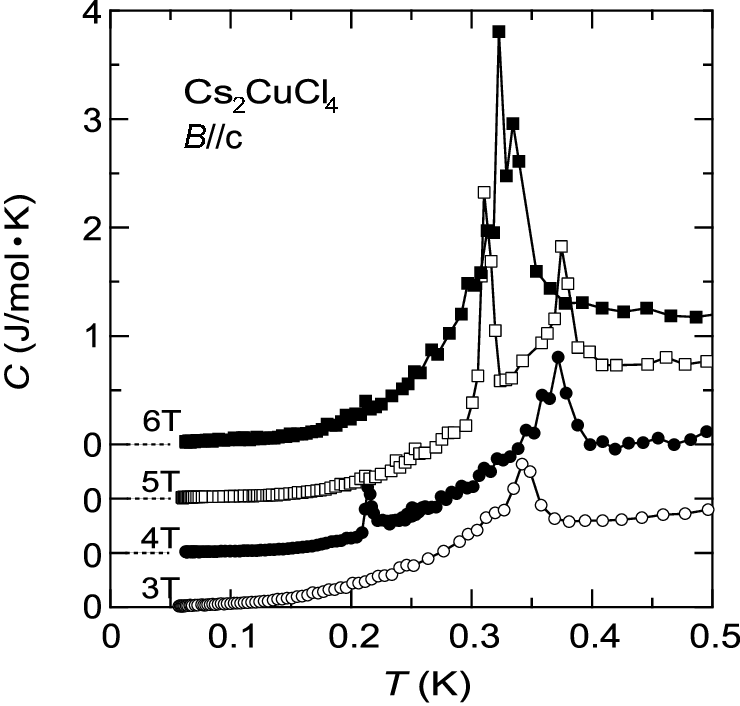}
\caption{\label{C_c} Specific heat as a function of temperature in magnetic fields along $c$-axis. Specific heat data in fields 4, 5 and 6\,T are
shifted upwards by 0.5, 1.0 and 1.5 J/mol$\cdot$K$^2$, respectively.}
\end{figure}

\begin{figure}[!ht]
\includegraphics[width=8cm,keepaspectratio]{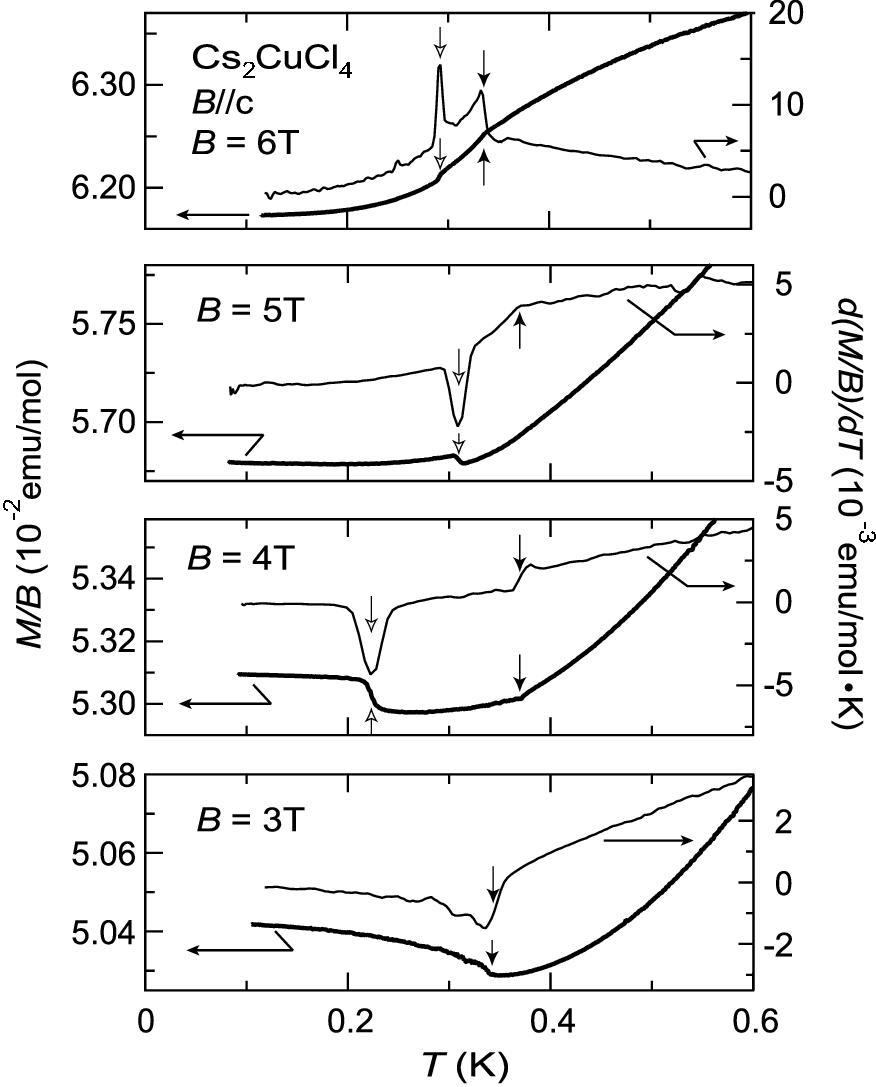}
\caption{\label{M(T)c}  Magnetization normalized by applied field $M/B$ (thick solid lines, left axis) and its derivative $d(M/B)/dT$ (thin solid
lines, right axis) as a function of temperature for $B\parallel c$. Vertical arrows indicate anomalies.}
\end{figure}

Complementary magnetization data vs. temperature for field along $b$ is shown in Fig.~\ref{M(T)b}. At 3 and 4\ T two anomalies are observed also in
$M/B(T)$ and its derivative $d(M/B)/dT$, at essentially the same temperatures as the peaks in specific heat, indicating that those two anomalies are
associated with magnetic phase transitions. The anomalies appear as kinks in $M/B(T)$ and steps in $d(M/B)/dT$. At 5\ T however the scaled
magnetization $M/B(T)$ only observes a clear anomaly at the lower of the two critical temperatures observed in specific heat. At 6\ T no anomaly is
visible in $M/B(T)$, but only the derivative $d(M/B)/dT$ shows a kink. The missing anomalies can however be seen in the raw capacitance data, plotted
in Fig.~\ref{torque}, which also contain information not only on the (longitudinal) magnetization but also the transverse spin components. At 4\,T
the capacitance in both zero and non-zero gradient field show two successive transitions indicated by solid and open arrows. Those are in good
agreement with the peaks observed in specific heat. Although there is no anomaly visible in the magnetization at 6\,T, the raw capacitance shows an
anomaly (see derivative of $d{\cal C}/dT$ in inset of Fig.~\ref{torque})) at the same temperature as the peak in specific heat. The capacitance in
non-zero gradient field contains information on the torque of the sample in addition to the magnetization, while that in zero gradient field does not
depend on magnetization but only on the torque. The torque contribution is subtracted by measuring the capacitance in zero gradient field (details of
measurement technique are described in Ref.~[5]). The fact that there is no anomaly in magnetization implies that subtraction of torque effect
cancels out the anomaly in the raw data. Therefore only the torque (transverse magnetization) has an anomaly and the longitudinal magnetization has
no anomaly at the critical temperatures for these missing anomalies.

For field along $c$ it has been reported from neutron scattering study that the spiral phase at zero field is suppressed by magnetic field of 1.4\,T
and above this field ordered spins form an incommensurate elliptical structure with elongation along the field direction~\cite{Coldea01}. The
elliptical phase is suppressed at 2.1\,T where the intensity of incommensurate magnetic Bragg peaks vanishes and the properties of the phase above
2.1\,T are still unknown. As shown in Fig.~\ref{M}(b), the suppression of the spiral phase is clearly seen as a step in magnetization (a sharp peak
in $dM/dB$) at 1.40\,T. In Fig.~\ref{AnormM(B)C} the magnetization at 0.07\,K and its derivative $dM/dB$ are expanded in order to show the four
anomalies above the spiral phase. In Fig.~\ref{AnormM(B)C}(a), $M(B)$ shows a small step (a peak in $\chi (B)$) at 2.05\,T, corresponding to the
suppression of the elliptical phase. As indicated by an open arrow in Fig.~\ref{AnormM(B)C}(a), a step in susceptibility at 2.18\,T is clearly seen,
indicating possibly a new phase which may exist only in a very small range of fields from 2.05 to 2.18\,T. The next transition occurs at 3.67\,T (for
increasing field) shown in Fig.~\ref{AnormM(B)C}(b). $M(B)$ has a step accompanied by a hysteresis, indicative of a first order transition.
Figure~\ref{AnormM(B)C}(c) shows another transition at 7.09\,T with a clear hysteresis.

Fig.\ \ref{C_c} shows specific heat in magnetic fields along the $c$-axis. At 3\,T only one transition is observed upon cooling, whereas at 4 and
5\,T two successive transitions are observed. The lower temperature transition is very sharp, related to the first order behavior (hysteresis) on
this transition line also observed in magnetization data $M(B)$ at 3.67\,T shown in Fig.~\ref{AnormM(B)C}(b). The lower temperature transition shifts
to higher temperatures with increasing field and almost merges with the upper transition at 6\,T.

Fig.\ \ref{M(T)c} shows complementary magnetization data vs. temperature. At 3\ T $M/B(T)$ and $d(M/B)/dT$ show a kink and a step at 0.35\,K,
respectively. At 4\,T the position of the kink (step in $d(M/B)/dT$) is shifted to slightly higher temperature and another step-like anomaly (a
negative peak in $d(M/B)/dT$) appears at lower temperatures 0.22\,K. At 5\,T this lower temperature step shifts to higher temperatures and the upper
temperature anomaly (kink) can not be seen in $M/B(T)$ but is manifested as a kink in $d(M/B)/dT$ at 0.38\,K. Again the anomaly is missing in
$M/B(T)$, but the raw capacitance data (not shown) exhibits an anomaly at 0.38\,K in good agreement with the specific heat result. As shown in the
top panel of Fig.\ \ref{M(T)c}, $M/B(T)$ and $d(M/B)/dT$ have two anomalies at 6\,T. Note that due to the first order character of the lower
temperature transition the anomalies of $M/B$ ($d(M/B)/dT$) indicated by open arrows in Fig.\ \ref{M(T)c} are steps (peaks) rather than kinks
(steps).

The phase diagrams for $B\parallel b$- and $c$-axis constructed using the anomalies discussed above are shown in Fig.~\ref{B-T}. The new data agree
with and complement earlier low-field neutron diffraction results (open triangles).\cite{Coldea01} Apart from the phase transition boundaries
identified above we have also marked the cross-over line between paramagnetic and antiferromagnetic short-range ordered(SRO) region, determined by
the location of the peak in the temperature dependence of the magnetization such as in Fig.\ \ref{chi}(a). The peak position $T_{\rm{max}}$ decreases
with increasing field and disappears above $B_c$, indicating suppression of antiferromagnetic correlations by magnetic field. For the field along $b$
and $c$-axis the phase diagrams are much more complicated than that for $B\parallel a$ which shows only one cone phase up to saturation
field~\cite{Coldea02,Rad05}. For $B\parallel b$ three new phases appear above the spiral phase. Two of these phases occupy small areas of the $B-T$
phase diagram. For $B\parallel c$ four new phases are observed in addition to the spiral and elliptical phases.

We note that the absence of an observable anomaly in the temperature dependence of the magnetization upon crossing the phase transitions near certain
fields (6\ T along $b$ and 5\ T along $c$) is consistent with Ehrenfest relation and is related to the fact that the transition boundary $T_c(B)$ is
near flat around those points. The relation between the shape of the phase boundary and the anomaly in $M(T)$ was discussed by T. Tayama, {\it et.
al.}\cite{Tay01} and is
\begin{equation}
\Delta\left(\frac{d M}{d
T}\right)=-\frac{dT_c}{dB}\Delta\left(\frac{C}{T}\right)
\label{Ehrenfest}
\end{equation}
where $\Delta(X)$ is the discontinuity of quantity $X$, $C$ is the specific heat and $T_c$ is the field-dependent critical temperature of second
order phase transition. This shows that the discontinuity in $dM/dT$ vanishes when $dT_c/dB=0$, i.e. when the phase boundary is flat in field. This
is indeed the case for 6\,T $\parallel b$ and at 5\,T $\parallel c$ [see Fig.\ \ref{B-T}], and here only a kink and no discontinuity is seen in
$dM/dT$.

\section{Conclusions}
We have studied the magnetic phase diagrams of Cs$_2$CuCl$_4$ by
measuring magnetization and specific heat at low temperatures and
high magnetic fields. The low-field susceptibility in the
temperature range from below the broad maximum to the Curie-Weiss
region is well-described by high-order series expansion
calculations for the partially frustrated triangular lattice with
$J'/J$=1/3 and $J$=0.385\,meV. The extracted ground state energy
in zero field obtained directly from integrating the magnetization
curve is nearly a factor of 2 lower compared to the classical
mean-field result. This indicates strong zero-point quantum
fluctuations in the ground state, captured in part by including
quantum fluctuations to order $1/S$ in a linear spin-wave
approach. The obtained $B-T$ phase diagrams for in-plane field ($B
\parallel b$ and $c$-axis) show several new intermediate-field
phases. The difference between the phase diagrams for $B\parallel
a, b$ and $c$ can not be explained by a semi-classical calculation
for the main Hamiltonian in Cs$_2$CuCl$_4$, of a frustrated 2D
Heisenberg model on an anisotropic triangular lattice with small
DM terms. Further neutron scattering experiments are needed to
clarify the magnetic properties of these new phases.

\begin{acknowledgements}
We would like to thank V. Yushankhai, D. Kovrizhin, D.A. Tennant,
M. Y. Veillette and J.T. Chalker for fruitful discussions, Z.
Weihong for sending the series data, P.~Gegenwart and
T.~L\"{u}hmann for technical support.
\end{acknowledgements}

\end{document}